\documentclass[submission,copyright,creativecommons]{eptcs}
\providecommand{\event}{ACL2 2018} 
\usepackage{underscore}           

\usepackage{graphicx}
\usepackage{amsmath}
\usepackage{amsfonts}
\usepackage{listings}
\title{The Fundamental Theorem of Algebra in ACL2}
\author{Ruben Gamboa \& John Cowles
\institute{University of Wyoming \\ Laramie, Wyoming}
\email{\{ruben,cowles\}@uwyo.edu}
}
\def\titlerunning{Fund. Thm. of Algebra}
\def\authorrunning{R. Gamboa \& J. Cowles}
\begin{document}
\maketitle

\begin{abstract}
We report on a verification of the Fundamental Theorem of Algebra in ACL2(r). 
The proof consists of four parts. First, continuity for both complex-valued 
and real-valued functions of complex numbers is defined, and it is shown 
that continuous functions from the complex to the real numbers achieve a minimum 
value over a closed square region. An important case of continuous real-valued, 
complex functions results from taking the traditional complex norm of a 
continuous complex function. We think of these continuous functions as having 
only one (complex) argument, but in ACL2(r) they appear as functions of two 
arguments. The extra argument is a ``context'', which is uninterpreted. 
For example, it could be other arguments that are held fixed, as in an 
exponential function which has a base and an exponent, either of which could 
be held fixed. Second, it is shown that complex polynomials are continuous, 
so the norm of a complex polynomial is a continuous real-valued function 
and it achieves its minimum over an arbitrary square region centered at the 
origin. This part of the proof benefits from the introduction of the ``context'' 
argument, and it illustrates an innovation that simplifies the proofs of 
classical properties with unbound parameters. Third, we derive lower and 
upper bounds on the norm of non-constant polynomials for inputs that are 
sufficiently far away from the origin. This means that a sufficiently 
large square can be found to guarantee that it contains the global minimum 
of the norm of the polynomial. Fourth, it is shown that if a given number 
is not a root of a non-constant polynomial, then it cannot be the global minimum. 
Finally, these results are combined to show that the global minimum must be a root 
of the polynomial. This result is part of a larger effort in the formalization of 
complex polynomials in ACL2(r).
\end{abstract}

\section{Introduction}

In this paper, we describe a verification of the Fundamental Theorem of Algebra
(FTA) in ACL2(r). That is, we prove that if $p$ is a non-constant, 
complex\footnote{By ``complex''
we mean the traditional mathematical view that the complex numbers are an extension
of the real numbers, not the ACL2 view that complex numbers are necessarily not
real.} polynomial with
complex coefficients, then there is some complex number $z$ such that $p(z)=0$.
The proof follows the outline of the first proof 
in~\cite{fine1997fundamental}, and it is a formal version of d'Alembert's proof
of 1746~\cite{gauss-proofs-fta}, which is illustrated (literally) 
in~\cite{visual-fta}.

\begin{figure}
\begin{center}
\includegraphics[scale=0.5]{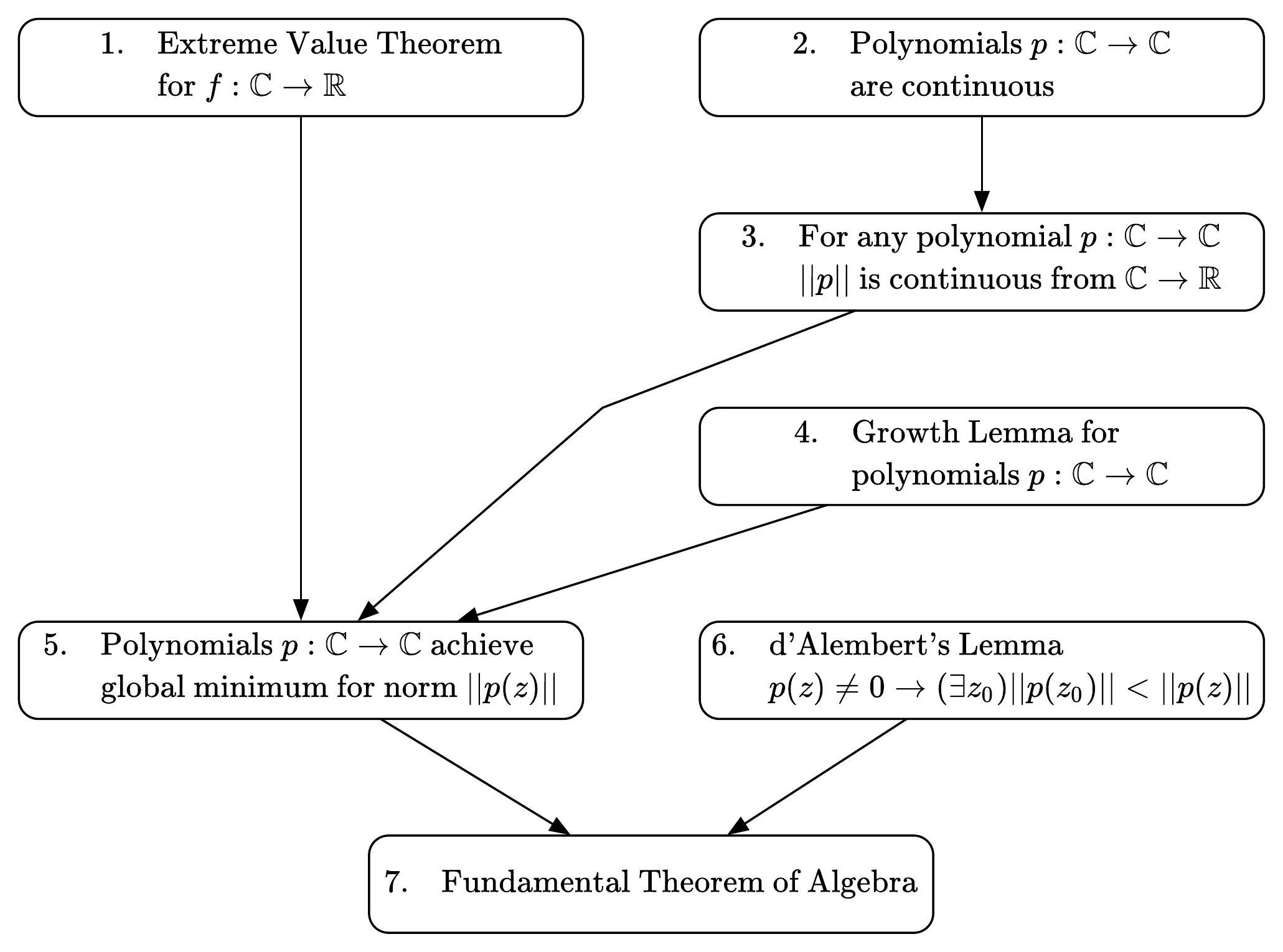}
\end{center}
\caption{Proof Outline}
\label{fig:proof-outline}
\end{figure}

Figure~\ref{fig:proof-outline} provides an outline of the proof, which comprises
three main strands. First (step 1 in Figure~\ref{fig:proof-outline}), we show that continuous functions from the complex
plane to the reals always achieve a minimum value in a closed, bounded, rectangular
region. The consequence to the FTA is that if $p$ is a complex polynomial, then
the function mapping $z\in\mathbb{C}$ to $||p(z)||$, where $||\cdot||$ denotes the
traditional complex norm, achieves a minimum value in
a square region centered at the origin (steps 2, 3, and 5). Second (step 4), we show that the norm of polynomials 
is dominated by the term of highest power. In particular, if 
$p(z) = a_nz^n + a_{n-1}z^{n-1} + \cdots + a_0$, then 
$\frac{1}{2}||a_n||\,||z||^n < ||p(z)|| < \frac{3}{2}||a_n||\,||z||^n$ for sufficiently
large $z$. These two facts can be combined to show that the global minimum of a
polynomial (norm) must be enclosed in a possibly large region around the origin, 
so the polynomial must achieve this global minimum. The third strand (step 6), known as
d'Alembert's Lemma, states that if $p$ is a complex polynomial and $z$ is such
that $p(z) \ne 0$, then there is some $z_0$ such that $||p(z_0)|| < ||p(z)||$.
This implies that the global minimum guaranteed by the previous lemmas must be
a root of the polynomial (step 7).

To our knowledge, this is the first proof of the Fundamental Theorem of Algebra
in the Boyer--Moore family of theorem provers, but it has been proved earlier in
other theorem provers, e.g., Mizar~\cite{Mil:FTA}, HOL~\cite{Har:FTA}, 
and Coq~\cite{Geu:FTA}.

\section{Continuity and the Extreme Value Theorem}

\subsection{Continuous Functions}

We begin the proof of the FTA with a proof of the Extreme Value Theorem for complex
functions. The first step is to define the notion of continuity for complex functions,
and this follows the pattern used before for real functions~\cite{GaKa:acl2r}. In
particular, we say that $f$ is continuous if for any standard number $z$, $f(z)$ is
close to $f(z_*)$ for any $z_*$ that is close to $z$. To be precise, two complex 
numbers are ``close'' if both the real and imaginary parts of their difference are 
infinitesimal. This also implies that the distance between the two points is
infinitesimal.

We also introduced in ACL2(r) the notion of a continuous function from the complex 
to the real numbers, and we proved that if both $f:\mathbb{C}\rightarrow\mathbb{C}$ 
and $g:\mathbb{C}\rightarrow\mathbb{R}$ are continuous, then so is $g\circ f:\mathbb{C}\rightarrow\mathbb{R}$. Moreover, we proved that the function given by the
traditional complex norm, $||a + bi|| = \sqrt{a^2+b^2}$, is continuous. So for any
continuous function $f:\mathbb{C}\rightarrow\mathbb{C}$, the function from $\mathbb{C}$
to $\mathbb{R}$ given by $h(z) = ||f(z)||$ is continuous.

An important difference from the development in~\cite{GaKa:acl2r} is that continuous
functions in this paper have two arguments in ACL2, even though we think of them as
functions of only one variable. This is similar to the way the function $x^n$ and
$a^x$ are introduced in elementary calculus. Those functions are thought of as functions
of the single variable $x$, and the derivatives are given as $nx^{n-1}$ and $a^x \ln{a}$,
even though both functions are simply slices of the bivariate function $x^y$.

In ACL2(r), we introduce the continuous function \texttt{ccfn} using \texttt{encapsulate}
as \texttt{(ccfn context z)}, where the argument \texttt{z} is the one that is allowed
to vary, whereas \texttt{context} is the argument that is held fixed. The non-standard
definition of continuity that we use here requires that close points be mapped to
close values, but only for standard \texttt{z}. Similarly, we require that close points are
mapped to close values only when both \texttt{z} and \texttt{context} are standard.

The \texttt{context} argument plays an important role, which can be illustrated by the
continuity of polynomials. Consider, for example, a proof that the function $x^n$ is
continuous. This would appear to be easy to prove by induction as follows:
\begin{enumerate}
	\item $x^0$ is a constant function, so it is definitely continuous.
	\item $x^n = x^{n-1} \cdot x$, and $x^{n-1}$ is continuous from the induction hypothesis,
and so is the product of continuous functions.
\end{enumerate}
However, this does not work. The formal definition of continuity uses the non-classical
notions of $close$: $standard(x) \wedge close(x, x_0) \rightarrow close(x^n, x_0^n)$.
But ACL2(r) restricts induction on non-classical functions to standard values of the
arguments. In this particular case, we could prove that $x^n$ is continuous, but only for
standard values of $n$, as was done in \cite{SaGa:sqrt}.

Setting that aside for now, we can imagine what would happen if we could establish 
(somehow) that complex polynomials are continuous. The next step in the proof would be
to show that since $p(z)$ is continuous in the complex plane, then $||p(z)||$ is continuous
from $\mathbb{C}$ to $\mathbb{R}$, and this could be done by functionally instantiating
the previous result about norms of continuous functions. Once more, however, we run into
a major complication. The problem is that we want to say this about \emph{all} polynomials, 
not just about a specific polynomial $p$. For example, we could use this approach to show
that the norm of the following polynomial is continuous:
\begin{lstlisting}
(defun p(z)
 (+ 1 (* #c(0 1) z) (* 3 z z)))
\end{lstlisting}
But that would lead to a proof that the polynomial $p$ has a root, not that all polynomials
have roots. To reason about all polynomials, we use a data structure that can represent any
polynomial and an evaluator function that can compute the value of a given polynomial at a
given point. For example, the polynomial above is represented with the list
\texttt{(1 \#c(0 1) 3)}, and the function \texttt{eval-polynomial} is defined so that
\texttt{(eval-polynomial '(1 \#c(0 1) 3) z) = (p z)}. But we cannot use functional
instantiation to show that the norm of \texttt{eval-polynomial} is continuous, because the
formal statement of continuity uses the non-classical notion of $close$. ACL2(r) restricts
functional instantiation so that pseudo-lambda expressions cannot be used in place of
functions when the theorem to be proved uses non-classical functions.

It should be noted that both of these restrictions are necessary! For example, $x^n$ is
``obviously'' continuous, but what happens when $n$ is large? Suppose that $x=1+\epsilon$,
where $\epsilon$ is small, so that $1$ and $1+\epsilon$ are close. From the binomial theorem,
$(1+\epsilon)^n = 1 + n\epsilon + \cdots$. The thing is that if $n$ is large, $n\epsilon$ is
not necessarily small. For instance, if $n = 1/\epsilon$, then $n\epsilon = 1$, and 
$(1+\epsilon)^n>2$, so it is not close to $1^n=1$. Yes, $x^n$ is continuous, but the 
non-standard definition of continuity only applies when $x$ \emph{and} $n$ are standard.

Similarly, the restriction for pseudo-lambda expressions is crucial. Consider, for example,
the theorem that $f(x)$ is standard when $x$ is standard. This is true, but we could not
use it to show that $x+y$ is standard when $x$ is standard by functionally 
instantiating $\lambda x.x+y$ for $f$. The proposed theorem is just not true when, for
example, $x=0$ and $y=\epsilon$.

So these restrictions are important for soundness, and we have to find a way to live with
them. In both cases, the key to doing so is the \texttt{context} argument. Consider
\texttt{eval-polynomial}, which has two arguments: the polynomial \texttt{poly} and the
point \texttt{z}. Here, \texttt{poly} plays the role of \texttt{context}. Notice how
\texttt{poly} is ``held fixed'' when we say that a polynomial is continuous over the
complex numbers. Because the \texttt{context} argument is used to refer to an arbitrary
polynomial, there is no need to use a pseudo-lambda
expression to introduce \texttt{poly}, so we can instantiate the theorem that
$||eval\text{-}polynomial(poly,z)||$ is continuous from $\mathbb{C}$ to $\mathbb{R}$
without running afoul of the restriction on free variables in functional instantiation
of non-classical formulas.

That leaves the continuity of polynomial functions to deal with. Here we want to show that
when $z$ is standard and close to $z_0$, $eval\text{-}polynomial(poly,z)$ is also close to
$eval\text{-}polynomial(poly,z_0)$. Again, the \texttt{context} argument plays a major role,
because the proof obligation of the \texttt{encapsulate} is actually weaker. It says that
$eval\text{-}polynomial(poly,z)$ must be close to $eval\text{-}polynomial(poly,z_0)$ only when both
$poly$ and $z$ are standard---and in that case we \emph{can} use induction to prove
the result. But notice that $poly$ is a list, so it is standard when it consists of
standard elements and has standard length, which is not the case for all polynomials.
As was the case before, this is the best we could expect. Consider, for instance, 
the polynomial $p(z) = Nz$ where $N$ is a large constant. Then $p(0)=0$ but
$p(\epsilon)=N\epsilon$ is not necessarily close to $0$.

But what about the Fundamental Theorem of Algebra? Does this mean that we can only prove
this theorem for standard arguments? Thankfully, that is not the case. We can proceed with
the proof of the Fundamental Theorem of Algebra for \emph{all} polynomials because of the
principle of transfer. This works because the statement of the Fundamental Theorem of Algebra 
is classical: if $p$ is a non-constant, complex polynomial with complex coefficients, 
then there is some complex number $z$ such that $p(z)=0$. Notice that statement does not
mention concepts like $standard$, $close$, or $small$. So if it holds for all standard
polynomials $poly$, it also holds for all polynomials, even the non-standard ones. The
same holds for other classical properties, such as the Extreme Value Theorem.

We believe that this approach is significantly cleaner than what we have done before.
For example, in~\cite{SaGa:sqrt} we used something similar to the context
argument, but the extra arguments were numbers. This means we were reasoning about 
functions of, say, 10 variables while holding 9 fixed, and if another fixed parameter 
was needed, we would have to redefine the constrained continuous function. In contrast,
the \texttt{context} parameter is only required to be standard, not necessarily a
number. So we could use it, for example, to hold a list of 9 fixed dimensions in one
setting, and 10 fixed dimensions in another one. The constrained function does not need
to change to accommodate the extra dimension, and nor do any of the generic theorems that 
were previously proved, e.g., the Extreme Value Theorem. As long as all the other dimensions
are fixed, they can be added to the \texttt{context}. 

A more traditional way of dealing with this problem is to realize that continuity itself is
a classical notion, so we can use a classical constrained function to introduce the notion
in ACL2(r). This is what we did in~\cite{CoGa:equivalences} where we explored equivalent
definitions of continuity and showed how versions of theorems like the Intermediate Value
Theorem could be proved for the classical and non-classical definitions of continuity.
But we also think that the approach used here works better, because there is no need to
have both classical and non-classical versions of all the theorems. Rather, the necessary
theorems can be proved using the non-classical definitions, which are simpler to use in
a theorem prover known for rewriting and induction, and then instantiating these (classical)
theorems when needed. That is exactly what we do here, where we prove the Extreme Value
Theorem using the intuitive notions of $close$, and then instantiate this theorem for all
polynomials using the fixed \texttt{context} parameter---without having to establish that
polynomials satisfy the $\epsilon$--$\delta$ definition of continuity.

\subsection{Proof of the Extreme Value Theorem}

\begin{figure}
\begin{center}
\includegraphics[scale=0.5]{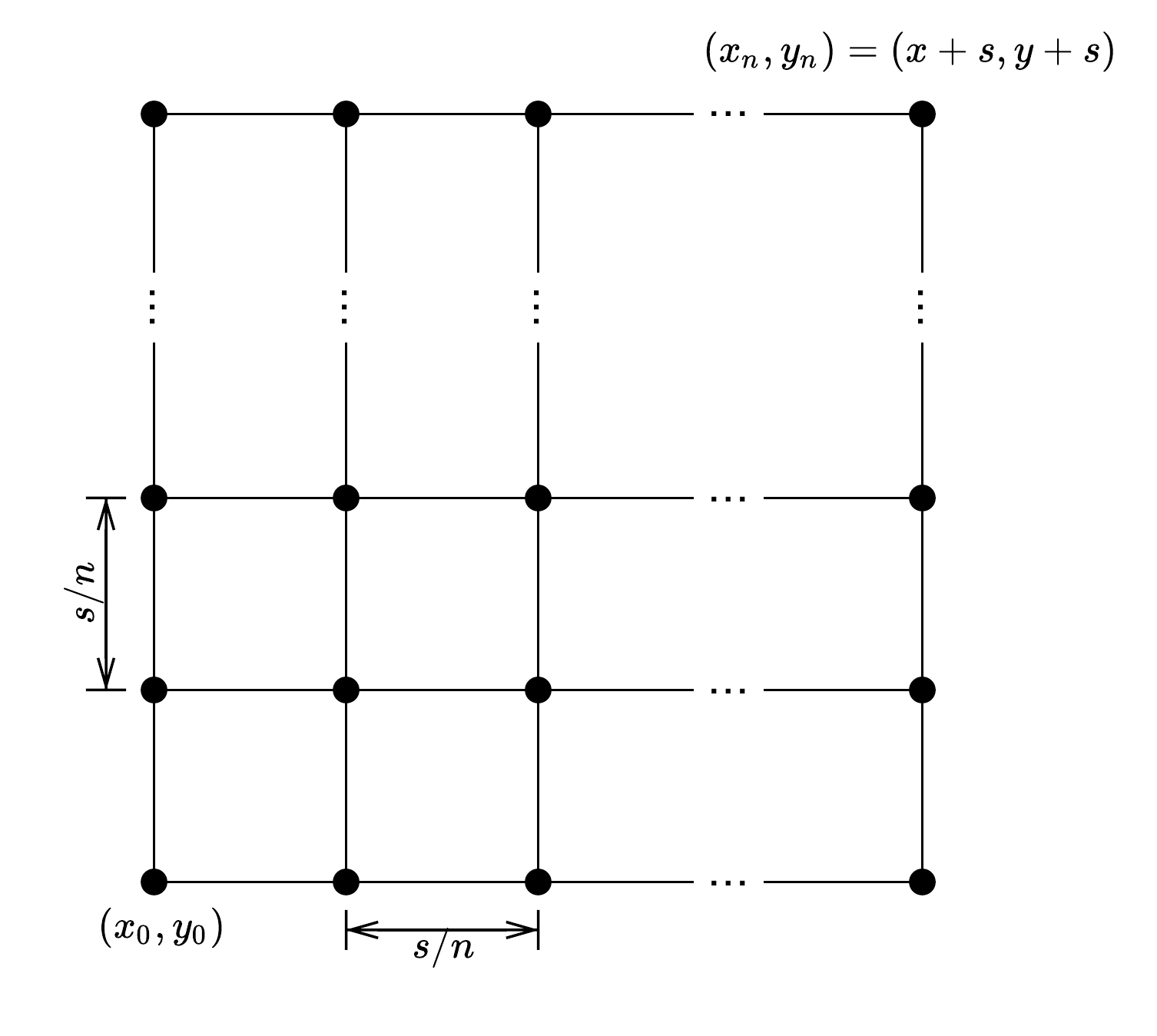}
\end{center}
\caption{Proof of Extreme Value Theorem}
\label{fig:evt-proof}
\end{figure}

We now turn attention to the proof of the Extreme Value Theorem for continuous functions
from the complex to the real numbers. The proof is similar in principle to the analogous
theorem for real functions, although it is more complicated since the relevant region is
a square instead of an interval. In both proofs, we use a constrained function to stand
for all possible continuous functions. In the earlier proof, we used \texttt{rcfn} to
stand for a \textbf{r}eal \textbf{c}ontinuous \textbf{f}u\textbf{n}ction. We also use
the name \texttt{ccfn} to stand for an arbitrary \textbf{c}omplex \textbf{c}ontinuous 
\textbf{f}u\textbf{n}ction. In this section, we will be using \texttt{crvcfn} which is a
constrained \textbf{c}omplex, \textbf{r}eal-\textbf{v}alued, \textbf{c}ontinuous 
\textbf{f}u\textbf{n}ction.

Figure~\ref{fig:evt-proof} shows the basic idea of the 
proof. The square of size $s\times s$ is subdivided into regions and the value of 
the function is found at the points in an $(n+1)\times(n+1)$ grid. Using recursion, it 
is straightforward to find the point on the grid where the function $f$ has a minimal 
value, say $z_m = (x_i, y_j)$.

At this point, we can assume that $(x_0, y_0)$ and $s$ are both standard, so the square
defined by the points $(x_0, y_0)$ and $(x_n, y_n)$ is also standard. This means that if
$z$ is inside the square, so is the standard part of $z$. In particular, the standard part
of any of the points in the grid must be inside the square region. And from this point on,
we can also assume that $n$ is a large number, so that adjacent points on the grid are a
distance at most $\sqrt{2}\,s/n$ apart, hence close to each other.

Since all the grid points are inside a standard square, they must be limited. This 
justifies using the non-standard definitional principle \texttt{defun-std} to define
the minimum as the standard part of $z_m$, written as ${}^* z_m$ or $standardpart(z_m)$. The non-standard
definitional principle assures us that the minimum is precisely ${}^* z_m$, but only
when $x_0$, $y_0$, and $s$ are standard.

Now, consider any standard point $z$ that lies inside the square region. It must be inside
one of the grid cells, and since adjacent grid points are close to each other, $z$ must 
also be close to the grid points in the enclosing cell, and since it is standard it must
be the standard part of those grid points---points that are close to each other have the
same standard part. So suppose that $z$ is close to $(x_k, y_l)$---as just observed, it
must be close to \emph{some} point on the grid. There are two cases to consider. If this
point happens to be equal to $z_m=(x_i, y_j)$, the minimum point on the grid, then as just
observed, it must be the case that its standard part ${}^*z_m$, must be equal to $z$. So
it follows trivially that $f({}^*z_m) \le f(z)$. Conversely, if $(x_k, y_l) \ne (x_i, y_j)
= z_m$, then it must be the case that $f(z_m) = f(x_i, y_j) \le f(x_k, y_l)$, because 
$z_m$ was chosen as the minimum point on the grid. Taking standard parts of both sides
shows that ${}^* f(z_m) = {}^* f(x_i, y_j) \le {}^* f(x_k, y_l)$. It is only necessary
to prove that ${}^* f(z_0) = f({}^* z_0)$ for all $z_0$, and again we have established
that $f({}^*z_m) \le f(z)$. So in either case, $f({}^* z_m)$ is at most the value of $f(z)$
for any standard $z$ in the square. Using the transfer principle with \texttt{defthm-std},
we generalized this statement to prove that ${}^* z_m$ is the minimum value over all
points in the square region from $(x_0, y_0)$ to $(x_0+s, y_0+s)$, for all values of
$x_0$, $y_0$, and $s$, not just the standard ones.

Although this proof is carried out for arbitrary continuous functions, the intent is 
to use functional instantiation to show that some specific function achieves its
minimum. Thus, it is very helpful to use quantifiers to state the Extreme Value Theorem,
so that a future use via functional instantiation does not need to recreate the 
functions used to search the grid, for example. The following definition captures
what it means for the point \texttt{zmin} to be the minimum point inside the square
or side length \texttt{s} with lower-left corner at \texttt{z0}:
\begin{lstlisting}
(defun-sk is-minimum-point-in-region (context zmin z0 s)
  (forall (z)
	  (implies (and (acl2-numberp z)
                  (acl2-numberp z0)
                  (realp s)
                  (< 0 s)
            			(inside-region-p 
                          z 
                          (cons (interval (realpart z0)
                                          (+ s (realpart z0)))
                                (interval (imagpart z0)
                                          (+ s (imagpart z0))))))
		         (<= (crvcfn context zmin) (crvcfn context z)))))
\end{lstlisting}
The Extreme Value Theorem, then, simply adds that there is some value of 
\texttt{zmin}
that makes this true. The existence is captured with the following definition:
\begin{lstlisting}
(defun-sk achieves-minimum-point-in-region (context z0 s)
  (exists (zmin)
          (implies (and (acl2-numberp z0)
                        (realp s)
                        (< 0 s))
                   (and (inside-region-p 
                                zmin 
                                (cons (interval (realpart z0)
                                                (+ s (realpart z0)))
                                      (interval (imagpart z0)
                                                (+ s (imagpart z0)))))
                        (is-minimum-point-in-region context 
                                                    zmin z0 s)))))

\end{lstlisting}
Using this definition, the Extreme Value Theorem can be stated simply as follows:
\begin{lstlisting}
(defthm minimum-point-in-region-theorem-sk
  (implies (and (acl2-numberp z0)
                (realp s)
                (< 0 s)
                (inside-region-p z0 (crvcfn-domain))
	            (inside-region-p (+ z0 (complex s s)) (crvcfn-domain)))
		   (achieves-minimum-point-in-region context z0 s))
  :hints ...)
\end{lstlisting}
It is then a simple matter to functionally instantiate this theorem for any
polynomial.

\section{Growth Lemma for Polynomials}

In the previous section, we proved the Extreme Value Theorem for continuous functions
from $\mathbb{C}$ to $\mathbb{R}$. In this section, we turn our attention to the values
$||p(z)||$ for polynomials $p$, and we are primarily interested in $||p(z)||$ when $z$
is sufficiently far from the origin.

Let $p(z)$ be given by
$p(z) = a_0 + a_1 z + a_2 z^2 + \cdots + a_n z^n$, where $a_n \ne 0$. It follows that
$$||p(z)|| = ||a_0 + a_1 z + a_2 z^2 + \cdots + a_n z^n||.$$
Using induction and the triangle inequality for $||\cdot||$, we proved that
$$||p(z)|| \le ||a_0|| + ||a_1 z|| + ||a_2 z^2|| + \cdots + ||a_n z^n||.$$
The norm $||\cdot||$ behaves nicely with products. In particular, we proved
that $||xy|| = ||x||\,||y||$, so
$$||p(z)|| \le ||a_0|| + ||a_1||\,||z|| + ||a_2||\,||z^2|| + \cdots + ||a_n||\,||z^n||.$$
Letting $A = \max ||a_i||$, it is easy to prove that
$$||p(z)|| \le A \left(||z^0|| + ||z^1|| + ||z^2|| + \cdots + ||z^n||\right).$$
We are only interested in the value of $||p(z)||$ when $z$ is sufficiently large, so
we can limit the discussion to those values of $z$ with $||z|| \ge 1$. A simple
induction shows that for those $z$, $||z|| \le ||z^n||$, so
$$||p(z)|| \le A \left(||z^n|| + ||z^n|| + ||z^n|| + \cdots + ||z^n||\right) = A (n+1) ||z^n||.$$
Another simple induction shows that $||z^n|| = ||z||^n$, so we have that
$$||p(z)|| \le A (n+1) ||z||^n.$$
Now, suppose that $\frac{A (n+1)}{K} \le ||z||$ for an arbitrary, real number 
$K$---clearly, for any fixed $K$, we can always find a $z$ large enough to make 
this true, since $A$ and $n$ are always fixed. Then $A (n+1) \le K||z||$, and we have
$$||p(z)|| \le A (n+1) ||z||^n \le K ||z||^{n+1}.$$

The last result holds for all polynomials $p$. Now we use that result to get both
lower and upper bounds for any polynomial $q$. In particular, let 
$q(z) = a_0 + a_1 z + a_2 z^2 + \cdots + a_n z^n$. Then $q$ can be written as 
$q(z) = (a_0 + a_1 z + a_2 z^2 + \cdots + a_{n-1} z^{n-1}) + a_n z^n.$ 
Applying the previous result to the polynomial in parentheses, we have that
$$q(z) \le K ||z||^n + ||a_n||\,||z||^n,$$
for any $K$ and $z$ such that $A n \le K||z||$. Choosing $K = ||a_n||/2$, we have 
that
$$q(z) \le K ||z||^n + ||a_n||\,||z||^n = \frac{||a_n||}{2} ||z||^n + ||a_n||\,||z||^n = \frac{3}{2} ||a_n||\,||z||^n.$$
That provides a nice upper bound for $||q(z)||$. In ACL2, this can be shown as
follows.
\begin{lstlisting}
(defthm upper-bound-for-norm-poly
  (implies (and (polynomial-p poly)
                (< 1 (len poly))
                (not (equal (leading-coeff poly) 0))
                (acl2-numberp z)
                (<= 1 (norm2 z))
                (<= (/ (* 2
                          (max-norm2-coeffs (all-but-last poly))
                          (1- (len poly)))
                       (norm2 (leading-coeff poly)))
                    (norm2 z)))
           (<= (norm2 (eval-polynomial poly z))
               (* 3/2
                  (norm2 (leading-coeff poly))
                  (expt (norm2 z) (1- (len poly))))))
  :hints ...)
\end{lstlisting}

We can find a lower bound on $||q(z)||$ by observing that
$$q(z) = a_n z^n - (-a_0 - a_1 z - a_2 z^2 + \cdots - a_{n-1} z^{n-1}),$$ 
and using the variant of the triangle inequality $||a-b|| \ge ||a| - ||b||$,
we have that
$$||q(z)|| \ge ||a_n z^n|| - ||-a_0 - a_1 z - a_2 z^2 + \cdots - a_{n-1} z^{n-1}||.$$ 
Further, since multiplying a value by $-1$ does not change its norm, we can write
this as 
$$||q(z)|| \ge ||a_n z^n|| - ||a_0 + a_1 z + a_2 z^2 + \cdots + a_{n-1} z^{n-1}||.$$
As we saw previously, 
$||a_0 + a_1 z + a_2 z^2 + \cdots + a_{n-1} z^{n-1}|| \le \frac{||a_n||}{2} ||z||^n$,
so
$$||q(z)|| \ge ||a_n z^n|| - ||a_0 + a_1 z + a_2 z^2 + \cdots + a_{n-1} z^{n-1}|| \ge ||a_n z^n|| - \frac{||a_n||}{2} ||z||^n = \frac{1}{2} ||a_n||\,||z||^n.$$
This can be shown in ACL2 as follows:
\begin{lstlisting}
(defthm lower-bound-for-norm-poly
  (implies (and (polynomial-p poly)
                (< 1 (len poly))
                (not (equal (leading-coeff poly) 0))
                (acl2-numberp z)
                (<= 1 (norm2 z))
                (<= (/ (* 2
                          (max-norm2-coeffs (all-but-last poly))
                          (1- (len poly)))
                       (norm2 (leading-coeff poly)))
                    (norm2 z)))
           (<= (* 1/2
                  (norm2 (leading-coeff poly))
                  (expt (norm2 z) (1- (len poly))))
               (norm2 (eval-polynomial poly z))))
  :hints ...)
\end{lstlisting}
Combining the previous two results, we have shown that
$$\frac{1}{2} ||a_n||\,||z||^n \le ||q(z)|| \le \frac{3}{2} ||a_n||\,||z||^n,$$
for values of $z$ with sufficiently large $||z||$.

Since the results hold for all $z$ with sufficiently large $||z||$, we can restrict
ourselves to $z$ such that $||z|| \ge 2\frac{||a_0||}{||a_n||}$. In this case, we have
that
$$q(z) \ge \frac{1}{2} ||a_n||\,||z||^n \ge \frac{1}{2} ||a_n||\,||z|| \ge \frac{1}{2} ||a_n|| \left(2\frac{||a_0||}{||a_n||}\right) = ||a_0||.$$
But observe that $q(0) = a_0$, so $||q(z)|| \ge ||q(0)||$.

\begin{figure}
\begin{center}
\includegraphics[scale=0.5]{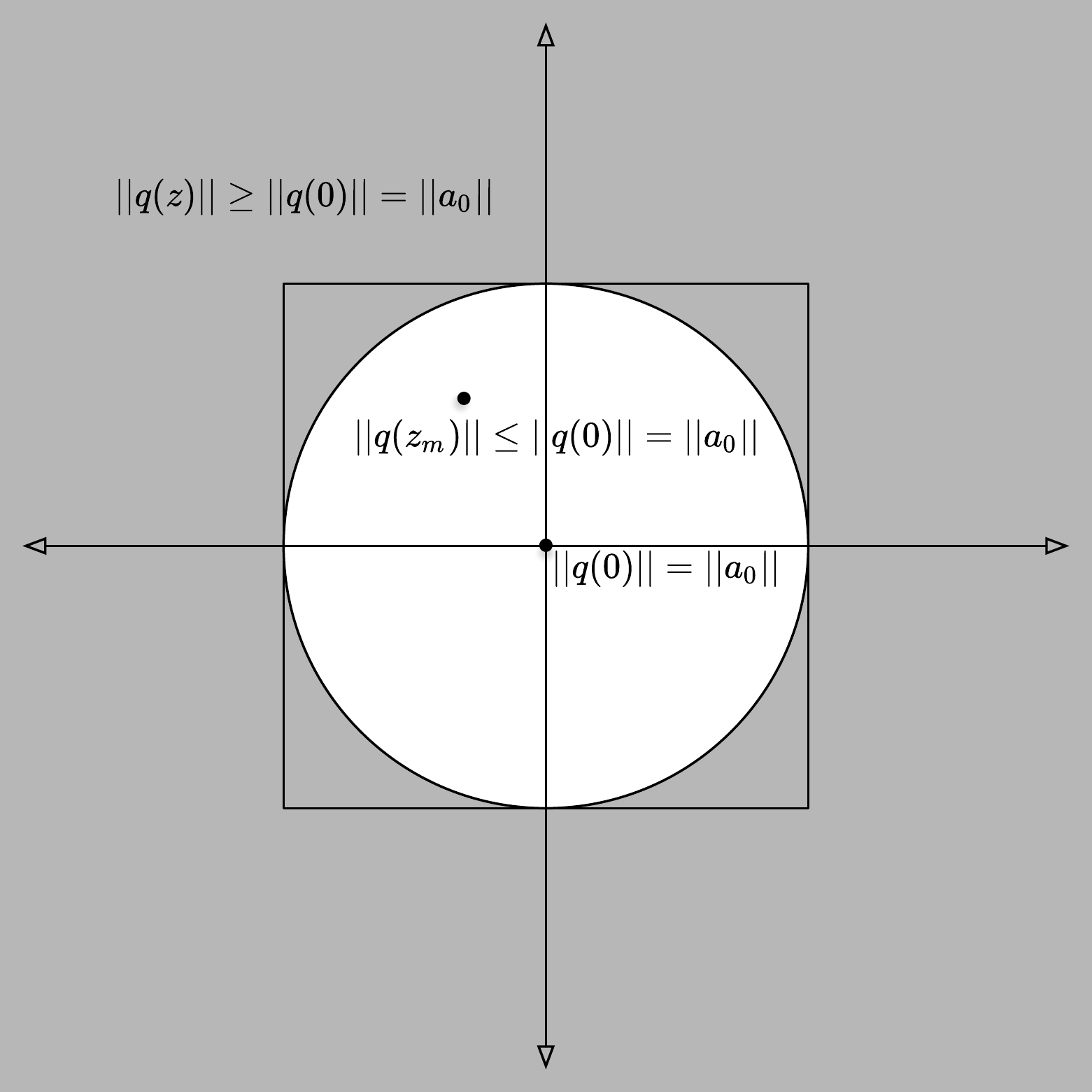}
\end{center}
\caption{Local Minimum Is a Global Minimum}
\label{fig:global-minimum}
\end{figure}

The situation is summarized in Figure~\ref{fig:global-minimum}. All points in the
gray area---that is, outside of the circle---are large enough that 
$||q(z)|| \ge ||a_0|| = ||q(0)||$. On the other hand, the minimum point $z_m$ guaranteed
by the Extreme Value Theorem lies inside the square, and since it is the minimum point
inside the square, it holds that $||q(z_m)|| \le ||q(0)|| = ||a_0||$. Combining these
two statements, we have that $||q(z_m)|| \le ||q(z)||$ for all $z$ whether they are
inside or outside of the circle. In other words, $z_m$ is not just a minimum inside the
square region. It is a \emph{global} minimum for $q$.

\section{d'Alembert's Lemma}

We now prove d'Alembert's Lemma, which states that for a non-constant polynomial $p$, if $z$ is such
that $p(z) \ne 0$, then there is some $z_0$ such that $||p(z_0)|| < ||p(z)||$.
In particular, if $p(z) \ne 0$ then $z$ cannot be a global minimum of $||p(\cdot)||$.

We prove this lemma by proceeding in a series of special cases. First, let's assume that
$a_0=1$ so that
$$p(z) = 1 + a_1 z + a_2 z^2 + \cdots + a_n z^n.$$
Let $k$ be the largest index such that $a_1 = a_2 = \cdots = a_{k-1} = 0$. We note that 
$k \le n$, since we assume throughout this paper that $a_n \ne 0$. Then,
$$
\begin{aligned}
p(z) &= 1 + a_k z^k + a_{k+1} z^{k+1} + \cdots + a_n z^n \\
	 &= 1 + a_k z^k + z^{k+1} q(z) \\
\end{aligned}
$$
where $q$ is a particular polynomial.
Of course, this means that
$$
\begin{aligned}
||p(z)|| &= ||1 + a_k z^k + z^{k+1} q(z)|| \\
         &\le ||1 + a_k z^k|| + ||z^{k+1} q(z)||
\end{aligned}
$$
Suppose that $s$ is real number such that $0 < s < 1$, and let
$z_s = \sqrt[k]{-\frac{s}{a_k}}$. We note in passing that the 
existence of such a $z_s$ is guaranteed by de~Moivre's lemma, 
which we proved as part of this effort, using the principal 
logarithmic function defined in~\cite{GaCo:inverses}.
\begin{lstlisting}
(defthm de-moivre-1
  (implies (and (acl2-numberp z)
                (natp n))
           (equal (expt z n)
                  (* (expt (radiuspart z) n)
                     (acl2-exp (* #c(0 1) n (anglepart z))))))
  :hints ...)
(defthm de-moivre-2
  (implies (and (acl2-numberp z)
                (posp n))
           (equal (expt (nth-root z n) n) z))
  :hints ...)
\end{lstlisting}
Since $z_s = \sqrt[k]{-\frac{s}{a_k}}$, we have that
$$a_k z_s^k = a_k \left(\sqrt[k]{-\frac{s}{a_k}}\right)^k = -s.$$
So $||1 + a_k z_s^k|| = ||1 - s|| = |1-s| = 1-s$, since $s$ is real and
$0<s<1$. This means that 
$$
\begin{aligned}
||p(z_s)|| &\le ||1 + a_k z_s^k|| + ||z_s^{k+1} q(z_s)|| \\
           &= 1 - s + ||z_s^{k+1} q(z_s)|| \\
           &= 1 - s + ||z_s||^k \, ||z_s|| \, ||q(z_s)|| \\
           &= 1 - s + \frac{s}{||a_k||} \, ||z_s|| \, ||q(z_s)|| \\
           &= 1 - s\left(1 - \frac{||z_s||}{||a_k||} ||q(z_s)||\right) \\
\end{aligned} 
$$
It's obvious that the term in parentheses is real and at most 1. What we want 
to show now is that it is actually positive and at most 1 for at least one
choice of $s$, and hence $z_s$.
That is, we will show that there is an $s$ such that
$\frac{||z_s||}{||a_k||} ||q(z_s)|| < 1$.

We find this $s$ by observing that whenever $z$ is such that $0 < ||z|| < 1$, then
$||q(z)|| \le M(n+1)$, where $M$ is the maximum norm of the coefficients,
i.e., $M = \max ||a_i||$. Let $r=\frac{||z||}{||a_k||} ||q(z)||$.
If $q(z) = 0$, then $r=0<1$. Otherwise, as long as $0<||z||<1$ and 
$0 < ||z|| < \frac{||a_k||}{M(n+1)}$, we have that $r = \frac{||z||}{||a_k||} ||q(z)||
< \frac{||a_k||}{||a_k||M(n+1)} ||q(z)|| \le \frac{1}{M(n+1)} M(n+1) = 1$.
So as long as $0<||z||<1$ and $0 < ||z|| < \frac{||a_k||}{M(n+1)}$, $0 \le r < 1$
(regardless of the value of $q(z)$.)
Therefore, 
$$
\begin{aligned}
||p(z)|| &\le 1 - s\left(1 - \frac{||z||}{||a_k||} ||q(z)||\right) \\
         &\le 1-s\left(1 - r\right) \\
         &< 1.
\end{aligned}
$$
The last inequality follows since $0 \le r < 1$, so $1-r>0$ and $s>0$. 
So it remains only 
to choose an $s$ such that $0 < ||z_s|| < \frac{||a_k||}{M(n+1)}$. Since
$z_s = \sqrt[k]{-\frac{s}{a_k}}$ this is equivalent to finding a positive
number $s$ small enough such that $s < \frac{||a_k||^{k+1}}{M^k(n+1)^k}$,
which is obviously possible since the right-hand side is positive. This is
summarized in the ACL2 theorem below:
\begin{lstlisting}
(defthm lowest-exponent-split-10
  (implies (and (polynomial-p poly)
                (equal (car poly) 1)
                (< 1 (len poly))
                (not (equal (leading-coeff poly) 0))
                (equal (car poly) 1)
                )
           (< (norm2 (eval-polynomial 
                           poly 
                           (fta-bound-1 poly 
                                        (input-with-smaller-value 
                                               poly))))
              1))
  :hints ...)
\end{lstlisting}
The function \texttt{fta-bound-1} corresponds to the choice of $x_s$, and
\texttt{input-with-smaller-value} finds a suitable value of $s$.

To complete the proof, observe that since $||p(z_s)|| < 1 = p(0)$, we have that 0 
cannot be the global minimum. It is worth remembering that the only thing we assumed
about this polynomial is that $a_0=p(0)=1$. We generalize this proof by removing this
assumption and letting $a_0\ne 1$. If it happens that $a_0=0$, then $p(0)=a_0=0$ and
the polynomial has a root. Otherwise, $a_0 \ne 0$ and we can define the new polynomial
$$p_1(z) = 1 + \frac{a_1}{a_0}z + \frac{a_2}{a_0}z^2 + \cdots + \frac{a_n}{a_0}z^n.$$
Notice that $p_1(z) = p(z)/a_0$, so $||p_1(z)|| = ||p(z)||/||a_0||$. This means that
if $0$ is not a global minimum of $||p||$, it cannot be a global minimum of $||p_1||$, 
either.

Finally, we remove the assumption that it is at the point $z=0$ that $p(z)\ne 0$.
Suppose, in fact, that $p(z_0)\ne 0$, and define
$$
\begin{aligned}
p_2(z) &= p(z+z_0) \\
	   &= a_0 + a_1(z+z_0) + a_2(z+z_0)^2 + \cdots + a_n(z+z_0)^n\\
	   &= b_0 + b_1 z + b_2 z^2 + \cdots + b_n z^n
\end{aligned}
$$
This is a more complicated transformation than the one defining $p_1$, but we showed
that $p_2$ is in fact a polynomial, that its highest exponent is $n$, and that
if $a_n\ne0$ then $b_n\ne0$ (in fact, $b_n=a_n$). That means that $p_2(0)=p(z_0)\ne0$,
and we can apply the previous theorem to show that $z_0$ is not a global minimum of
$||p_2||$.

There is one remaining assumption. Throughout this paper, we have been assuming that
the leading coefficient $a_n \ne 0$. But what about a polynomial such as
$$p(z) = a_0 + a_1 z + a_2 z^2 + \cdots + a_{n-1} z^{n-1} + 0 z^n$$
To generalize the results to this polynomial, we construct a new polynomial
$trunc(p)$ that simply removes all terms $a_i$ where $a_i=a_{i+1}=\cdots=a_n=0$.
It is easy to show that $trunc(p)$ is a polynomial and that it agrees with $p$ at
all values of $z$. Moreover, we say that a polynomial $p$ is not constant if at
least one of $a_1$, $a_2$, \dots, $a_n$ is not equal to 0. Then it is easy to 
prove that $trunc(p)$ is also a non-constant polynomial, such that its leading 
coefficient is not zero. That means that the theorems above apply to $trunc(p)$, 
and thus to $p$.

This completes the proof of d'Alembert's Lemma: if $p$ is a polynomial and $z$ is 
such that $p(z) \ne 0$, then $z$ is not a global minimum of $p$. It is then 
trivial to prove the Fundamental Theorem of Algebra. Since we already know that 
there is a $z_m$ such that $||p(\cdot)||$ has a minimum at $z_m$, it must be the 
case that $p(z_m)=0$.

In ACL2, we can write the condition that a given polynomial has a root using the
following Skolem definition:
\begin{lstlisting}
(defun-sk polynomial-has-a-root (poly)
  (exists (z)
          (equal (eval-polynomial poly z) 0)))
\end{lstlisting}
The Fundamental Theorem of Algebra can then be stated as follows:
\begin{lstlisting}
(defthm fundamental-theorem-of-algebra-sk
  (implies (and (polynomial-p poly)
                (not (constant-polynomial-p poly)))
           (polynomial-has-a-root poly))
  :hints ...)
\end{lstlisting}

\section{Conclusion}

We have shown a proof in ACL2(r) of the Fundamental Theorem of Algebra. 
Table~\ref{tab:proof-stats} gives an indication of the size of the proof
in ACL2(r).
The proof
follows the 1746 proof of d'Alembert, which was maligned by Gauss on the grounds 
that it was not sufficiently rigorous. Gauss was right, in that it took another 
hundred years for the notion of continuity to be sufficiently rigorous to prove 
the Extreme Value Theorem, on which the proof depends. Gauss offered a solution 
to d'Alembert's dilemma, in the form of a more formal proof of the Fundamental 
Theorem of Algebra, which nevertheless suffered from its own lack of rigor with 
respect to algebraic curves. Aware of this, Gauss proceeded to offer three other 
proofs of the Fundamental Theorem of Algebra, all essentially correct. We are in 
the midst of studying complex polynomials in ACL2, and Gauss's proofs offer a 
fertile playground for this purpose, so we expect to formalize some of those 
proofs in ACL2(r) in the future.

\begin{table}
\begin{center}
\begin{tabular}{l|p{2.5in}|l|l}
\textbf{File} & \textbf{Description} & \textbf{\#Definitions} & \textbf{\#Theorems} \\[1.5em]
\hline
\texttt{norm2} & \raggedright Basic facts about the complex norm & 2 & 72 \\[1.5em]
\texttt{complex-lemmas} & \raggedright Basic facts from complex analysis & 0 & 25 \\[1.5em]
\texttt{de-moivre} & \raggedright deMoivre's theorem & 2 & 36 \\[1.5em]
\texttt{complex-continuity} & \raggedright Extreme value theorem for complex functions & 19 & 96 \\[1.5em]
\texttt{complex-polynomials} & \raggedright Polynomials are continuous, achieve a minimum in a closed area, and have arbitrarily large values for large enough arguments; d'Alembert's Lemma; and Fundamental Theorem of Algebra & 41 & 197 \\[2.5em]
Total & & 64 & 426 \\ 
\end{tabular}
\end{center}
\caption{Proof Statistics}
\label{tab:proof-stats}
\end{table}

\nocite{*}
\bibliographystyle{eptcs}
\bibliography{fta}

\begin{thebibliography}{10}
\providecommand{\bibitemdeclare}[2]{}
\providecommand{\surnamestart}{}
\providecommand{\surnameend}{}
\providecommand{\urlprefix}{Available at }
\providecommand{\url}[1]{\texttt{#1}}
\providecommand{\href}[2]{\texttt{#2}}
\providecommand{\urlalt}[2]{\href{#1}{#2}}
\providecommand{\doi}[1]{doi:\urlalt{http://dx.doi.org/#1}{#1}}
\providecommand{\bibinfo}[2]{#2}

\bibitemdeclare{article}{gauss-proofs-fta}
\bibitem{gauss-proofs-fta}
\bibinfo{author}{Harel \surnamestart Cain\surnameend} (\bibinfo{year}{2018}):
  \emph{\bibinfo{title}{C. F. Gauss's Proofs of the Fundamental Theorem of
  Algebra}}.
\newblock \urlprefix\url{http://math.huji.ac.il/~ehud/MH/Gauss-HarelCain.pdf}.

\bibitemdeclare{inproceedings}{CoGa:equivalences}
\bibitem{CoGa:equivalences}
\bibinfo{author}{J.~\surnamestart Cowles\surnameend} \&
  \bibinfo{author}{R.~\surnamestart Gamboa\surnameend} (\bibinfo{year}{2014}):
  \emph{\bibinfo{title}{Equivalence of the Traditional and Non-Standard
  Definitions of Concepts from Real Analysis}}.
\newblock In: {\sl \bibinfo{booktitle}{Proceedings of the 12th International
  Workshop of the ACL2 Theorem Prover and its Applications}},
  \doi{10.1007/3-540-36126-X_17}.

\bibitemdeclare{book}{fine1997fundamental}
\bibitem{fine1997fundamental}
\bibinfo{author}{B.~\surnamestart Fine\surnameend} \&
  \bibinfo{author}{G.~\surnamestart Rosenberger\surnameend}
  (\bibinfo{year}{1997}): \emph{\bibinfo{title}{The Fundamental Theorem of
  Algebra}}.
\newblock \bibinfo{series}{Undergraduate Texts in Mathematics},
  \bibinfo{publisher}{Springer New York}, \doi{10.1007/978-1-4612-1928-6}.

\bibitemdeclare{misc}{wiki-fta}
\bibitem{wiki-fta}
\bibinfo{author}{\surnamestart {Fundamental theorem of algebra}\surnameend}
  (\bibinfo{year}{2001}): \emph{\bibinfo{title}{Fundamental theorem of algebra
  --- {W}ikipedia{,} The Free Encyclopedia}}.
\newblock
  \urlprefix\url{https://en.wikipedia.org/wiki/Fundamental_theorem_of_algebra}.

\bibitemdeclare{inproceedings}{GaCo:inverses}
\bibitem{GaCo:inverses}
\bibinfo{author}{R.~\surnamestart Gamboa\surnameend} \&
  \bibinfo{author}{J.~\surnamestart Cowles\surnameend} (\bibinfo{year}{2009}):
  \emph{\bibinfo{title}{Inverse Functions in {ACL2(r)}}}.
\newblock In: {\sl \bibinfo{booktitle}{Proceedings of the Eighth International
  Workshop of the ACL2 Theorem Prover and its Applications (ACL2-2009)}},
  \doi{10.1145/1637837.1637846}.

\bibitemdeclare{article}{GaKa:acl2r}
\bibitem{GaKa:acl2r}
\bibinfo{author}{R.~\surnamestart Gamboa\surnameend} \&
  \bibinfo{author}{M.~\surnamestart Kaufmann\surnameend}
  (\bibinfo{year}{2001}): \emph{\bibinfo{title}{Nonstandard analysis in
  {ACL2}}}.
\newblock {\sl \bibinfo{journal}{Journal of Automated Reasoning}}
  \bibinfo{volume}{27}(\bibinfo{number}{4}), pp. \bibinfo{pages}{323--351},
  \doi{10.1023/A:1011908113514}.

\bibitemdeclare{misc}{Geu:FTA}
\bibitem{Geu:FTA}
\bibinfo{author}{H.~\surnamestart Geuvers\surnameend},
  \bibinfo{author}{F.~\surnamestart Wiedijk\surnameend},
  \bibinfo{author}{J.~\surnamestart Zwanenburg\surnameend},
  \bibinfo{author}{R.~\surnamestart Pollack\surnameend} \&
  \bibinfo{author}{Ha~\surnamestart Barendregt\surnameend}:
  \emph{\bibinfo{title}{The ``Fundamental Theorem of Algebra'' Project}}.
\newblock \urlprefix\url{http://www.cs.kun.nl/~freek/fta/index.html}.

\bibitemdeclare{inproceedings}{Har:FTA}
\bibitem{Har:FTA}
\bibinfo{author}{J.~\surnamestart Harrison\surnameend} (\bibinfo{year}{2001}):
  \emph{\bibinfo{title}{Complex Quantifier Elimination in {HOL}}}.
\newblock In: {\sl \bibinfo{booktitle}{TPHOLs 2001: Supplemental Proceedings}},
  pp. \bibinfo{pages}{159--174}.
\newblock
  \urlprefix\url{http://www.inf.ed.ac.uk/publications/online/0046/b159.pdf}.

\bibitemdeclare{inproceedings}{Mil:FTA}
\bibitem{Mil:FTA}
\bibinfo{author}{R.~\surnamestart Milewski\surnameend} (\bibinfo{year}{2000}):
  \emph{\bibinfo{title}{Fundamental Theorem of Algebra}}.
\newblock In: {\sl \bibinfo{booktitle}{Journal of Formalized Mathematics}},
  \bibinfo{volume}{12}.

\bibitemdeclare{inproceedings}{SaGa:sqrt}
\bibitem{SaGa:sqrt}
\bibinfo{author}{J.~\surnamestart Sawada\surnameend} \&
  \bibinfo{author}{R.~\surnamestart Gamboa\surnameend} (\bibinfo{year}{2002}):
  \emph{\bibinfo{title}{Mechanical Verification of a Square Root Algorithm
  using {T}aylor's Theorem}}.
\newblock In: {\sl \bibinfo{booktitle}{Formal Methods in Computer-Aided Design
  (FMCAD'02)}}, \doi{10.1007/3-540-36126-X_17}.

\bibitemdeclare{article}{visual-fta}
\bibitem{visual-fta}
\bibinfo{author}{Daniel~J. \surnamestart Velleman\surnameend}
  (\bibinfo{year}{2015}): \emph{\bibinfo{title}{The Fundamental Theorem of
  Algebra: A Visual Approach}}.
\newblock {\sl \bibinfo{journal}{The Mathematical Intelligencer}}
  \bibinfo{volume}{37}(\bibinfo{number}{4}), pp. \bibinfo{pages}{12--21},
  \doi{10.1007/s00283-015-9572-7}.

\end{thebibliography}
\end{document}